\documentclass[12pt]{iopart}

\usepackage{textcomp}

\usepackage{graphicx}

\usepackage[caption=false]{subfig}

\usepackage{fixltx2e}

\expandafter\let\csname equation*\endcsname\relax
\expandafter\let\csname endequation*\endcsname\relax

\usepackage[cmex10]{amsmath}
\usepackage[separate-uncertainty=true]{siunitx}

\usepackage{color}
\newif\ifcmtr
\cmtrfalse
\ifcmtr
\newcommand{\cmtr}[1]{[\color{red} \textbf{#1} \normalcolor]}
\else
 \newcommand{\cmtr}[1]{ %
}
\fi

\usepackage{xspace} 
\def\SiN{SiRN\xspace}
\def\SiO{SiO$_2$\xspace}

\usepackage{nicefrac}

\usepackage{siunitx}

\usepackage[utf8]{inputenc}

\begin{document}

\title{Elastocapillary folding using stop-programmable hinges fabricated by 3D micro-machining}

\author{A. Legrain $^1$, J.W. Berenschot $^1$, N.R. Tas $^1$ and L. Abelmann $^{1,2}$}
\address{$^1$ MESA+ Institute for Nanotechnology, University of Twente, Enschede, The Netherlands}
\address{$^2$ KIST Europe, Saarbrücken, Germany}

\ead{a.b.h.legrain@utwente.nl}

\date{\today}

\begin{abstract}

We show elasto-capillary folding of silicon nitride objects with accurate folding angles between flaps of  \SI{70.6+-0.1}{\degree} and demonstrate the feasibility of such accurate micro-assembly with a final folding angle of \SI{90}{\degree}.  The folding angle is defined by stop-programmable hinges that are fabricated starting from silicon molds employing accurate three-dimensional corner lithography.  This nano-patterning method exploits the conformal deposition and the subsequent timed isotropic etching of a thin film in a 3D shaped silicon template.  The technique leaves a residue of the thin film in sharp concave corners which can be used as an inversion mask in subsequent steps.  Hinges designed to stop the folding at  \SI{70.6}{\degree} were fabricated batchwise by machining the V-grooves obtained by KOH etching in (110) silicon wafers;  \SI{90}{\degree} stop-programmable hinges were obtained starting from silicon molds obtained by dry etching on (100) wafers.  The presented technique is applicable to any folding angle and opens a new route towards creating structures with increased complexity, which will ultimately lead to a novel method for device fabrication.

\end{abstract}

\maketitle
 
\section{Introduction}

\subsection{Self-folding of 3D micro-structures}

The fabrication of 3D micro-structures has become an important field of interest in the scientific community over the past three decades~\cite{Leong2010}.
Traditional mask-based approaches, such as photo-lithography and its developments (including X-ray lithography, electron-beam lithography and dip-pen nanolithography), have proven to be inadequate for fabricating truly 3D-patterned structures.  The main limitations include: an inherent two-dimensionality, size limitations, being time-consuming, and demanding a complex fabrication~\cite{Madou1997}.

Fabrication examples of 3D structures abound in nature.  Salt crystallization and the folding of  protein or DNA are processes that engineers dream of reproducing in a laboratory with as much precision and reproducibility as is seen in nature.  The process by which disordered components are organized into patterns or structures without human intervention is known as   ``self-assembly'' or, by analogy with the previously mentioned top--down methods, a   ``bottom--up'' approach~\cite{Whitesides2002}.  Although great proofs-of-concept have been published~\cite{Ariga2008,Mastrangeli2009}, such engineering suffers from a too high level of uncertainty, as pointed out by Gracias \emph{et al.} in their excellent review~\cite{Leong2010}.  Therefore, they prefer the use of a more deterministic form of self-assembly known as ``self-folding'' or ``micro-origami''.  Combining the strengths of both lithography and self-assembly, the final 3D structure is predetermined by the linkages between the different parts that are assembled.  The obvious link with origami, the ancient Japanese art of folding paper, provided its name to this technique~\cite{Brittain2001}.  While origami-like planar structures are fabricated using standard micro-machining techniques, many methods of self-folding have been investigated, some more common than others.  These methods include ultrasonic pulse impact~\cite{Kaajakari2003}, pneumatics~\cite{Lu2006}, electroactive swelling~\cite{Guan2005,Kim2005c}, thermal actuation of polymer films~\cite{Luo2006,Stoychev2011,Ionov2011,Liu2011}, thin-film stress-based assembly (TFSA)~\cite{Arora2006,Schmidt2001,Chua2003,Leong2008,Cho2010}, magnetic forces~\cite{Judy1997,Boncheva2005,Iwase2006,Nichol2006,Gagler2008}, and capillary forces.  Surface tension is probably the most common method of self-folding.  In the micro/nanometer world, interfacial forces dominate over body forces such as gravity, making them a perfect candidate for the self-folding of micro-structures.  Syms was the first to introduce this method by using solder pads which are melted to power assembly before solidification in their final state~\cite{Syms1993,Syms2000}.  More recently, this method has had great nanoscale applications as a result of the work of Gracias \emph{et al.}~\cite{Leong2007,Filipiak2009,Cho2009a}.
For a complete overview of self-folding techniques, see the recent reviews~\cite{Leong2010,Randall2011a,Shenoy2012,Ionov2013}.

An elegant macro-scale illustration of self-folding by surface-tension
is Bico~\emph{et al.}~\cite{Py2007,Py2009,Roman2010},
who demonstrated the spontaneous wrapping of thin millimeter-sized polydimethylsiloxane (PDMS) sheets
around a water droplet: the so-called elastocapillary folding technique.  Using the same concept, our group has demonstrated
the fabrication of silicon nitride 3D micro-objects by capillary forces
in which the actuating liquid, in our case water, disappears as a
result of its spontaneous evaporation.
Final closure is assured by the strong cohesion between flaps without
the need for solder, and the assembly is carried out under ambient conditions
either by simply depositing water on top of the structures~\cite{vanHonschoten2010} or by providing a liquid through a tube at the centre of the objects~\cite{Legrain2014}.

A crucial feature of self-folded objects is to pre-determine their final shape.  Some techniques require a locking mechanism, for example, some research on self-folding by magnetic interaction~\cite{Judy1997,Iwase2006}, while another treatment of magnetic self-folding relies on plastic deformation~\cite{Gagler2008}.  Using solder assembly, the quantity of melting material determines the final folding angle~~\cite{Syms1993,Leong2007,Syms2003}.  Likewise, in TFSA, the final radius of curvature is a function of the stresses in the different layers, with curvatures ranging from a few millimeters down to nanometers~\cite{Arora2006,Schmidt2001,Chua2003}.  The final shape from using self-folding polymer films can be controlled by designing several small shrinking hinges in series~\cite{Stoychev2011,Ionov2011,Ionov2013}.

Structures folded by elasto-capillary interactions are limited in terms of their final three-dimensional shapes.  Folding ceases once the moving flaps encounter an obstacle, typically a nearby flap, and the elasticity of the hinges causes the objects to re-open if not enough stiction is present between the flaps~\cite{Py2007,vanHonschoten2010,Legrain2014}.  The work presented here aims at extending the scope of elastocapillary folding of silicon nitride micro-objects by predefining the final assembly.  The principle of these stop-programmable hinges is presented in Fig.~\ref{fig:Principle}.

 Made of a thick rigid part and a thin flexible part, these complex hinges are designed in such a way that the final assembly angle is predefined by their shape.  After folding through evaporation of water, the two opposite thick parts meet and adhere.  The final folding angle therefore depends on the initial angle between the substrate and the thick parts of these smart hinges.

\begin{figure}
\centering
\includegraphics[width=.8\linewidth]{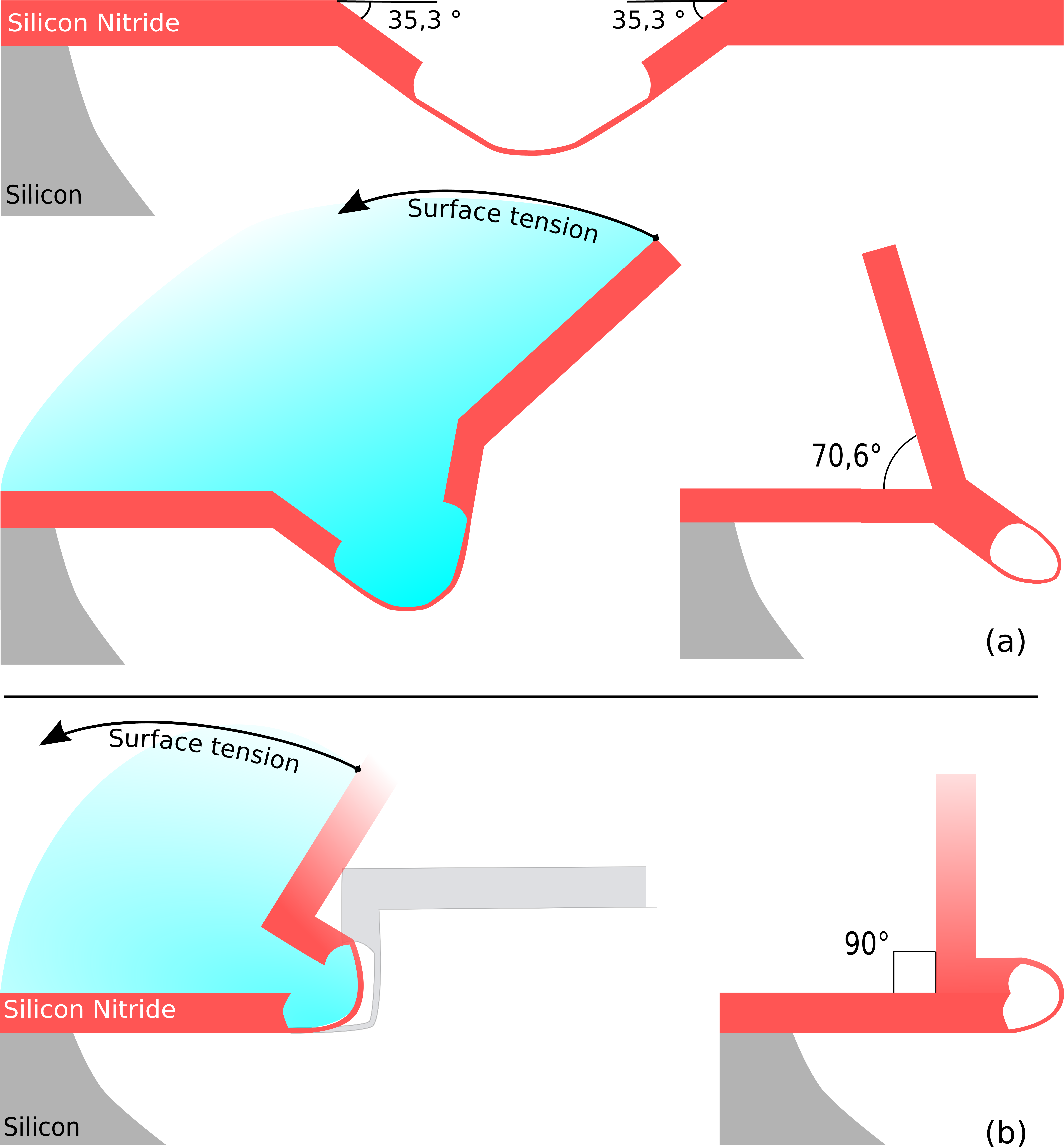}
\caption{The stop-programmable folding principle.
The design of the complex hinges is such that once folded, the flap forms a predefined angle with the planar support.  Self-folding of the structure is enabled through evaporation of water and decrease
of the liquid--air interface of the meniscus.  (a) \SI{70.6}{\degree} stop-programmable hinge.   (b): \SI{90}{\degree} stop-programmable hinge.   In both cases, the flaps adhere thanks to a sufficiently large stiction area and there is no need for a locking mechanism.}
\label{fig:Principle_sph}
\end{figure}

\subsection{Corner lithography and self-folding}

\begin{figure}
\centering
\includegraphics[width=.8\linewidth]{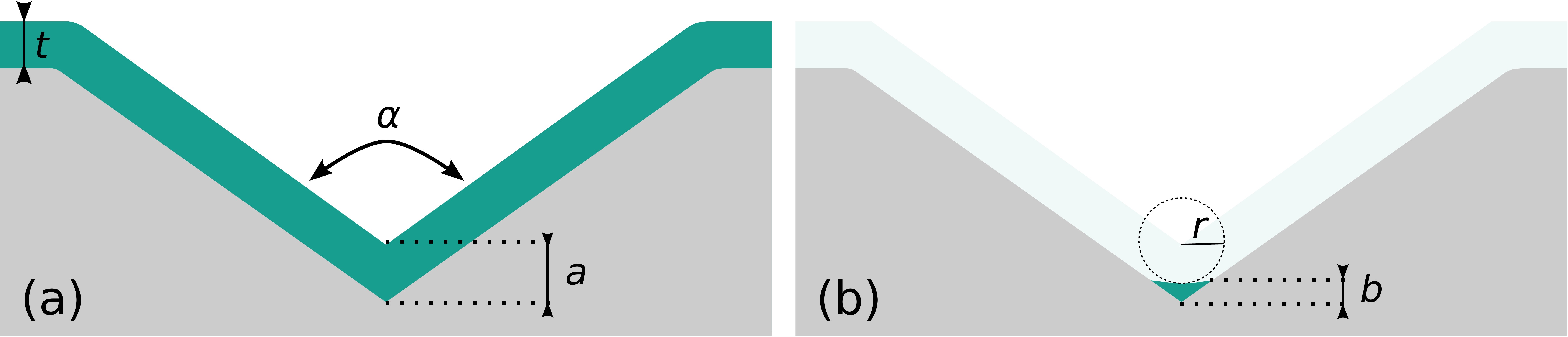}
\caption{Corner lithography in a sharp corner.  (a): When a conformal layer of thickness $t$ is deposited over a concave corner of opening $\alpha$, the effective thickness of material at the corner is $a=\nicefrac{t}{\sin (\nicefrac{\alpha}{2})}> t$\cite{Sarajlic2005e}.  (b): After isotropic etching by an amount of $r$, material with thickness $b=a-r$ remains in the corner.}
\label{fig:Principle corner litho}
\end{figure}

Corner lithography is a wafer scale nano-patterning technique that offers the
opportunity to form structures in sharp concave corners, independently of their orientation in space.  The conformal deposition of a material layer over a patterned substrate will result in a greater effective thickness in any sharp concave corner.  Isotropic etching therefore yields nano-features as presented in Fig.~\ref{fig:Principle corner litho}.  
This technique was first developed and used in our laboratory to create silicon nitride nano-wire pyramids~\cite{Sarajlic2005e}.  We then extended the scope of this technique by demonstrating the use of the structures formed by corner lithography as a mask for subsequent patterning steps~\cite{Berenschot2008}.  In the meantime, Yu \emph{et al.} demonstrated the fabrication of nano-ring particles and photonic crystals using corner lithography~\cite{Yu2009}.  More recently, our group has continued the development of this technique and demonstrated the parallel nano-fabrication of fluidic components with cell culturing application~\cite{Berenschot2012}, as well as the wafer-scale fabrication of nanoapertures~\cite{Burouni2013} and the machining of silicon nitride 3D fractal structures~\cite{Berenschot2013}.  

\begin{figure}
\centering
\includegraphics[width=.8\linewidth]{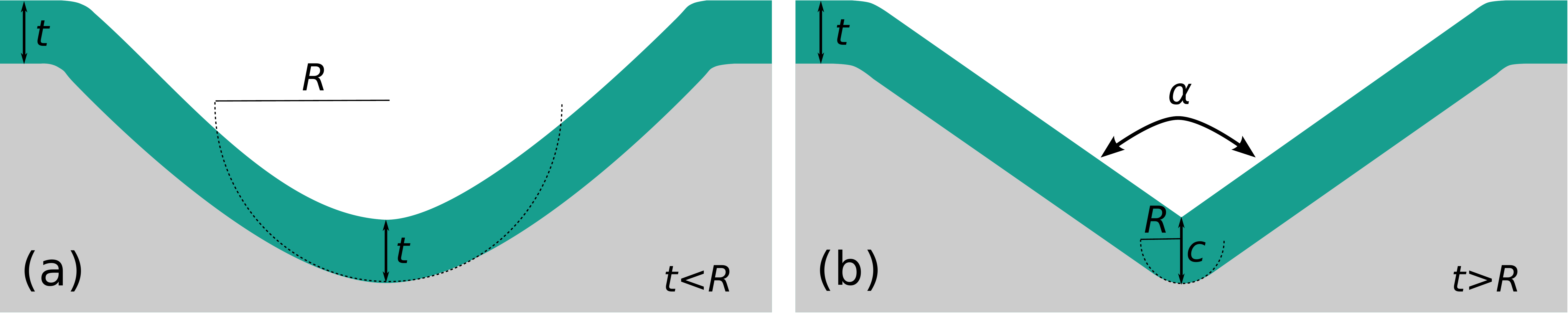}
\caption{Corner lithography in a rounded corner.  (a): When a material layer of thickness $t$ is conformally deposited over a mold with radius of curvature $R$  that is greater than $t$, the thickness at the tip is unchanged.  (b): On the other hand, when $R$ is less than $t$, a concave corner is created and the effective thickness is $c=R+\nicefrac{(t-R)}{\sin (\nicefrac{\alpha}{2})}$.}
\label{fig:Design_criterion}
\end{figure}

In this paper, we use corner lithography to fabricate the smart hinges presented in Fig.~\ref{fig:Principle}.  Sharp features must be avoided when it comes to bending or folding, since they lead to an extreme concentration of stress~\cite{Kondo1984,Pilkey2008}.  Consequently, corner lithography needs to be performed in rounded molds for our purposes.  This situation leads to conditions on the radius of curvature of the mold, as well as on the thicknesses of the subsequently deposited material.  In general, the total thickness of the materials that are deposited in rounded molds must be greater than their radius of curvature so as to have some material remaining after the isotropic etching, as shown in Fig.~\ref{fig:Design_criterion}.

\section{Fabrication process flow}

\cmtr{Here confusion between making the molds and patterning the nitride.}

In this section we present the fabrication steps necessary to build the stop-programmable hinges depicted in Fig.~\ref{fig:Principle_sph}.  The main steps are the same for the fabrication of both folding angles, although  \SI{90}{\degree} stop-programmable hinges require the use of wet etching to pattern the vertical sidewalls, unlike the \SI{70.6}{\degree} complex hinges, for which everything can be accessed by dry etching.  The two following sections extensively describe the process flow, with the first part describing the whole procedure for \SI{70.6}{\degree} smart hinges, and the second part pointing out the differences when fabricating \SI{90}{\degree} stop-programmable hinges.

\subsection{\SI{70.6}{\degree} stop-programmable hinges}

\begin{figure}
\centering
\includegraphics[width=1\linewidth]{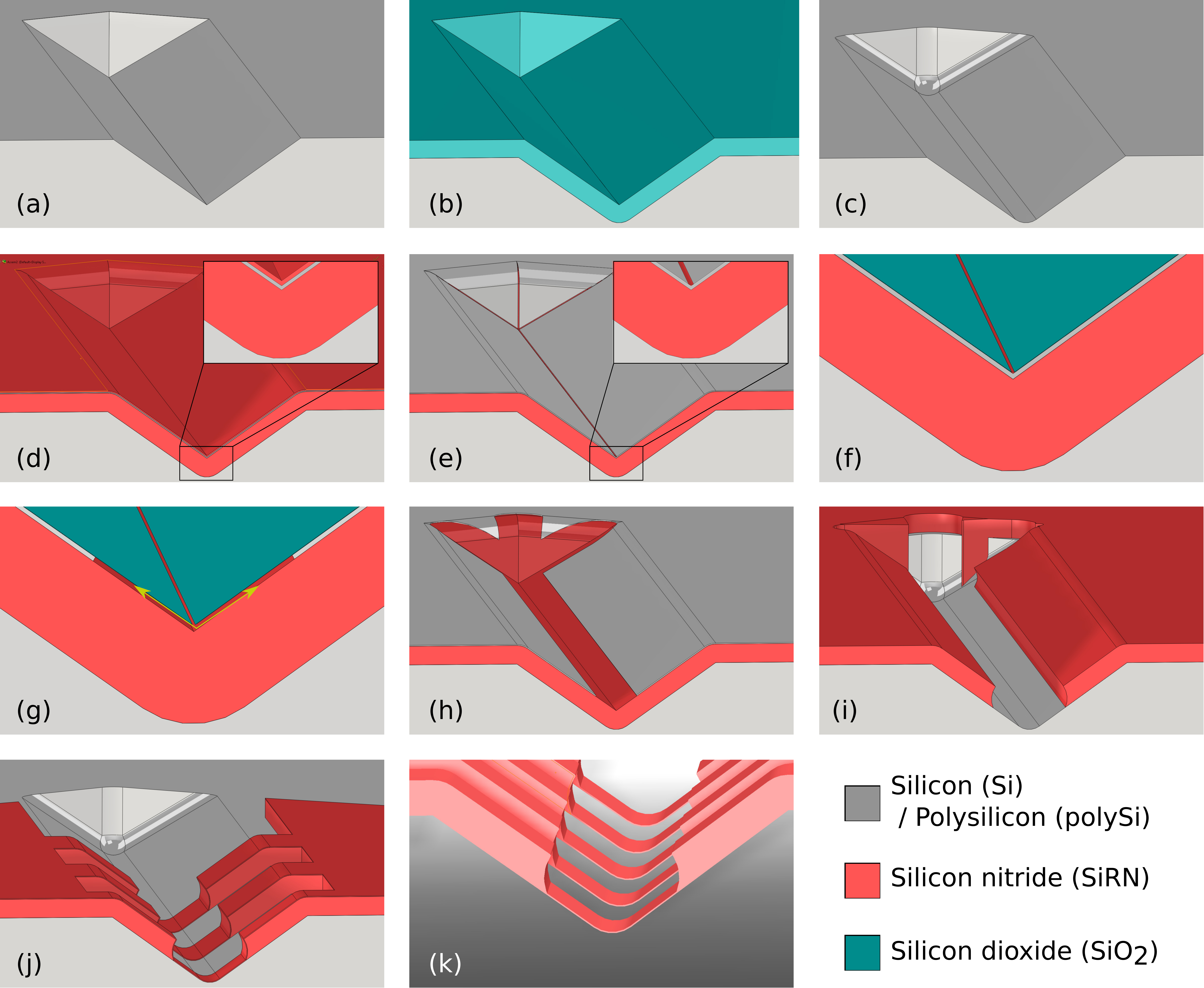}
\caption{Fabrication of a stop-programmable hinge, example of a \SI{70.6}{\degree} smart hinge.  (a):~On a (110) oriented silicon wafer, V-grooves are etched.  (b)~and~(c):~These grooves are rounded off by means of oxidation and subsequent etching.  (d):~A \SiN/polySi/\SiN stack of layers is conformally deposited.  (e):~The top  \SiN layer is isotropically wet etched using a time stop such that  material remains in the corner.  \SiN wires run in all three directions.  (f):~This remaining \SiN nanowire is used as a protection mask during partial oxidation of the underlying polySi layer.  (g):~After removal of the \SiN line the polySi layer is retracted.  (h):~\SiO is stripped.  (i):~Using the polySi layer as a mask, an opening is etched in the \SiN  layer.  An oxidation and subsequent wet etching step follow to remove the polySi layer (not shown here).  (j):~After deposition of flexible layer of \SiN, a mask is applied through lithography  and the overall geometry of the structure is determined by directive ion etching.  (k): A last lithography step follows for protection of the \SiN objects during semi-isotropic etching of Si.  Once released from the substrate, the smart hinge is ready to be self-folded.}
\label{fig:Fab Outline corner litho}
\end{figure}

The strategy is similar to that presented by our lab in previous publications~\cite{Berenschot2008,Berenschot2012}.  Corner lithography is here employed to create a masking layer which will be subsequently used to etch the underlying body layer before being removed.  Fig.~\ref{fig:Fab Outline corner litho} shows the procedure step by step for machining a \SI{70.6}{\degree} stop-programmable hinge.

The initial Si (silicon) molds will define the shape of our final object.  The opening angle $\alpha$ of the molds, see Fig.~\ref{fig:Principle corner litho} and~\ref{fig:Design_criterion}, defines the final folding angle, $\beta$, through the relation $\beta=\pi-\alpha.$  Silicon has a face-centred cubic structure with a well-defined lattice.  The angle between the top $<$110$>$ plane and the $\{$111$\}$
planes is exactly \SI{35.3}{\degree}, as depicted in Fig.~\ref{fig:Principle}.  Etching the molds with KOH on (110) oriented wafers yields well-defined openings with the desired opening angle ($\alpha$=\SI{109.4}{\degree}, $\beta$=\SI{70.6}{\degree}) with sharp transitions between the different planes, see Fig.~\ref{fig:Fab Outline corner litho}~(a).  Note that the mask pattern, consisting of rectangular openings, needs to be rotated by 54.7\si{\degree} with respect to the vertical $<$111$>$ planes in order to get rectangular  V-grooves with KOH etchant.

 The bending of sharp objects is to be avoided since it leads to extreme stress concentrations~\cite{Kondo1984,Pilkey2008}.  We therefore use an oxidation step to round off the molds, Fig.~\ref{fig:Fab Outline corner litho}~(b).  Kim \emph{et al.} showed that high temperature oxidation is an efficient method to round off sharp silicon V-grooves~\cite{Kim2002}.  Linear relations between the oxidation time and the final achieved radius of curvature were experimentally found by the authors for (100) oriented wafers but not for (110) wafers.  Our own oxidation experiments followed by SEM inspections allowed us
 to determine a similar relation for a 1150\si{\degree}C wet oxidation step applied to (110) wafers:
 
 \begin{equation}
 R=\SI{124(72)}+(\SI{94(7)}{}) \sqrt{t}
 \end{equation}
  with $R$ the final radius of curvature in~\si{\nm} and $t$ the oxidation time in \si{\minute}.  A \SI{98}{\minute} oxidation step yields a \SI{1}{\um} radius of curvature and is used for our fabrication, Fig.~\ref{fig:Fab Outline corner litho}~(b).  It is difficult to use a higher radius of curvature since that would imply a deposition of a thicker material layer, as described in Fig.~\ref{fig:Design_criterion}~(a).  Once \SiO (silicon dioxide) is stripped in HF, the molds are ready for corner lithography.
 
First, two layers are deposited by low-pressure chemical vapor deposition (LPCVD): first, a thick silicon rich nitride (\SiN)  layer that is to be structured to form the thick part of the smart hinges, see Fig.~\ref{fig:Principle}, followed by a polysilicon (polySi) layer.  The total thickness of material here needs to be greater than the initial radius of curvature $R$ as was emphasized in Fig.~\ref{fig:Design_criterion}.  Typically, we deposit a \SI{1}{\um} \SiN layer and a \SI{150}{\nm} polySi layer in a mold where the radius of curvature $R$ is \SI{1}{\um}.  In any case, the following design criterion should be respected:
 
 \begin{equation}  \label{eq:design_criterion}
  t_\text{\SiN-1}+t_\text{polySi} \ge R
  \end{equation}
where $t_\text{\SiN-1}$ and $t_\text{polySi}$ stand for the thickness of the bottom \SiN and the polySi layers, respectively.
 
 On top of this stack, a last conformal layer of \SiN  is deposited.  Provided that the design criterion has been respected, the cross section of the stack should look like Fig.~\ref{fig:Fab Outline corner litho}~(d).  There are no constraints on the thickness of the top \SiN layer, but it  needs to be perfectly known since this layer will be time-etched in phosphoric acid ($H_{3}PO_{4}$) to form \SiN nanowires at the bottom of the grooves as shown in Fig.~\ref{fig:Fab Outline corner litho}~(e).  In the case of a $\alpha= \SI{109.4}{\degree}$ opening, the material is \SI{22}{\percent} thicker at the tip (see Fig.~\ref{fig:Principle corner litho}) and over-etching of the \SiN layer is allowed within this limit.  We have used a thin \SiN layer of about \SI{100}{nm} to reduce the etching time.
 
 Thanks to the newly formed nanowires running in the three dimensions, the underlying polySi layer can be partially oxidized (Fig.~\ref{fig:Fab Outline corner litho}~(f)) using \SiN nanowires as an inversion mask.  Half of the polySi thickness can be consumed during this step without issues.  Once the nanowires are selectively etched, the polySi layer can be retracted starting from the tips where nicely defined access points are now formed, step (g).  TMAH as an etchant is a nice option here since its selectivity between \SiN and \SiO is high and the etching speed not too fast to be controllable, but KOH could also be used.  Careful timing is necessary during the retraction, since the length of the flexible parts of the smart hinges are determined during this step, see Fig.~\ref{fig:Principle_sph}.  Typically, a TMAH solution at \SI{25}{wt\percent} at \SI{95}{\degree}C attacks polySi at a speed of  \SI{1}{\um\per\minute}.
 
The above steps result in a patterned polySi layer on top of the thick \SiN layer, Fig.~\ref{fig:Fab Outline corner litho}~(h).  Using the polySi layer as a mask, it is now possible to etch the underlying \SiN layer in HF~\SI{50}{\percent}.  Due to the isotropic nature of the etchant, a retraction equal to the thickness of the material etched will occur under the polySi masking layer.  Moreover, when the thickness of the \SiN layer to be etched is greater than the initial radius of curvature $R$, an over-etching is necessary to remove the surplus material (Fig.~\ref{fig:Design_criterion}), its thickness can be calculated by 
 
  \begin{equation}
   c=\frac{(t-R)}{\sin (\frac{\alpha}{2})}
   \label{eq:surplus_material}
   \end{equation}
 with $t$ the thickness of the conformal layer and $\alpha$ the opening angle of the mold.

 In order to remove the polySi masking layer, an oxidation step should be preferred over a wet etching step since we want to conserve the rounding of the mold.  This short oxidation step will consume the polySi material and slightly increase the radius of curvature of the mold.  Moreover, the oxidation of the  Si substrate will yield a specific shape, known as a bird's beak, at the transition between the Si and the \SiN~\cite{Shankoff1980}.  This shape is not visible in Fig.~\ref{fig:Fab Outline corner litho} but will be shown in the results part of the paper (Fig.~\ref{fig:706_last_steps}).
 
After removal of the polySi layer, Fig.~\ref{fig:Fab Outline corner litho}~(i), the flexible part of the hinges is deposited by LPCVD.  The bending stiffness of the hinges  is $B=\nicefrac{Et^3}{12(1-\nu^2)}$ for thin plates, where $E$ is Young's modulus and $\nu$ is Poisson's ratio~\cite{landau1986}).  It is highly dependent on the thickness $t$ of the thin plate, and the thinner is this layer, the more flexible is the hinge.  \SI{100}{\nm} thin hinges offer both mechanical solidity and flexibility~\cite{vanHonschoten2010,Legrain2014}.  On the other hand, the final release of the foldable objects requires the etching of the Si molds in SF$_6$, which also slightly attacks \SiN.  The thinning of the hinges during the release step should therefore be taken into account when depositing the thin \SiN layer.  We measured a selectivity of around 1000 between silicon and \SiN on the part of our etching system (Adixen AMS100 Reactive Ion Etcher).  
 
 A second lithography step follows to define the overall geometry of the smart hinges.  Making holes along the length of the hinges permits reducing their stiffness and facilitates the folding.  Lithography in molds up to \SI{10}{\um} deep is relatively straightforward when using an appropriate photoresist.  In this specific case, dry etching can be used to remove the \SiN, step (j).  An extra lithography step must be performed to protect the \SiN structures by photoresist during the semi-isotropic etching of the silicon.  Once released from the silicon substrate, the stop-programmable hinges are ready for assembly.  When designing the masks, extra care was taken to assure that all structures were etched free from the substrate at the same time during the last step.  Ideally, the mask openings should be of the same size and placed at the same distance from the stop-etching point.
 
\begin{figure}
\centering
\includegraphics[width=.8\linewidth]{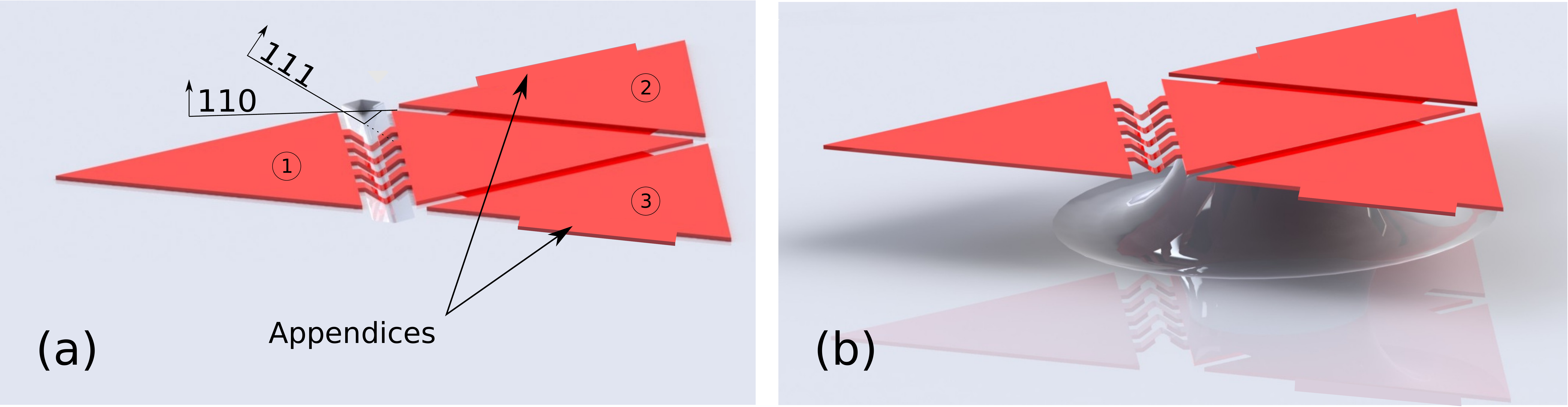}
\caption{Fabrication of a tetrahedron folding pattern, extra steps.  (a): Since only one of the hinges lies on the correct intersection of the planes, standard lithography is used to define flat hinges for the other faces.  Faces \raisebox{.5pt}{\textcircled{\raisebox{-.9pt} {2}}} and \raisebox{.5pt}{\textcircled{\raisebox{-.9pt} {3}}}  are designed with small appendices on their sides to allow them to lock onto face \raisebox{.5pt}{\textcircled{\raisebox{-.9pt} {1}}} while folding.  (b): Under-etching of Si by semi-isotropic etching of silicon (SF\textsubscript{6} etchant).  Etching is stopped when the hinges are free and the central flap rests on a silicon pillar.}
\label{fig:pattern_tetra}
\end{figure}

The \SI{70.6}{\degree} stop-programmable hinges thus fabricated can be used to self-fold perfectly defined tetrahedrons.  Since only one out of three hinges in a tetrahedron pattern lie at the right intersection between the planes in the silicon lattice,  as shown in Fig.~\ref{fig:pattern_tetra}, only one smart hinge can be formed in the way that was just described.  The other two hinges are therefore flat junctures made by standard micro-machining.  An extra lithography step followed by dry etching is therefore necessary between steps (i) and (j) in Fig.~\ref{fig:Fab Outline corner litho}.

\subsection{\SI{90}{\degree} stop-programmable hinges}
\label{sec:90deg_fab}

The procedure to fabricate \SI{90}{\degree}  stop-programmable hinges is nearly the same as for the \SI{70.6}{\degree} hinges described above, except for two important differences: making \SI{90}{\degree} molds in silicon is very difficult using wet etching---our attempts using correctly oriented (110) wafers always resulted in tiny bumps at the bottom edges of the molds---and the vertical \SiN sidewalls obtained in these molds cannot be patterned using directive dry etching.  Steps (a) and (i--j) in Fig.~\ref{fig:Fab Outline corner litho} therefore differ when it comes to the micromachining of \SI{90}{\degree} stop-programmable hinges.

Given that the depth of the molds is small, cryogenic dry etching is a good option for our purpose.  Unlike the BOSCH processes, cryogenic etching yields smooth sidewalls~\cite{Jansen2009}, which are crucial for our sensitive corner-lithography technique.  Retraction of the mask is also a well known problem in dry etching and might be an issue for corner-lithography.  In general, when developing a dry etching step for the purpose of performing corner lithography later on inside the molds, any concave corners other than the ones at the bottom of the molds should be avoided.  This includes potential roughness of the masking material and retraction of the mask.  Moreover, the final opening angle $\alpha$ is highly dependent on the etching conditions (type of mask used, loading, gas flows) and requires precise tuning.

\begin{figure}
\centering
\includegraphics[width=1\linewidth]{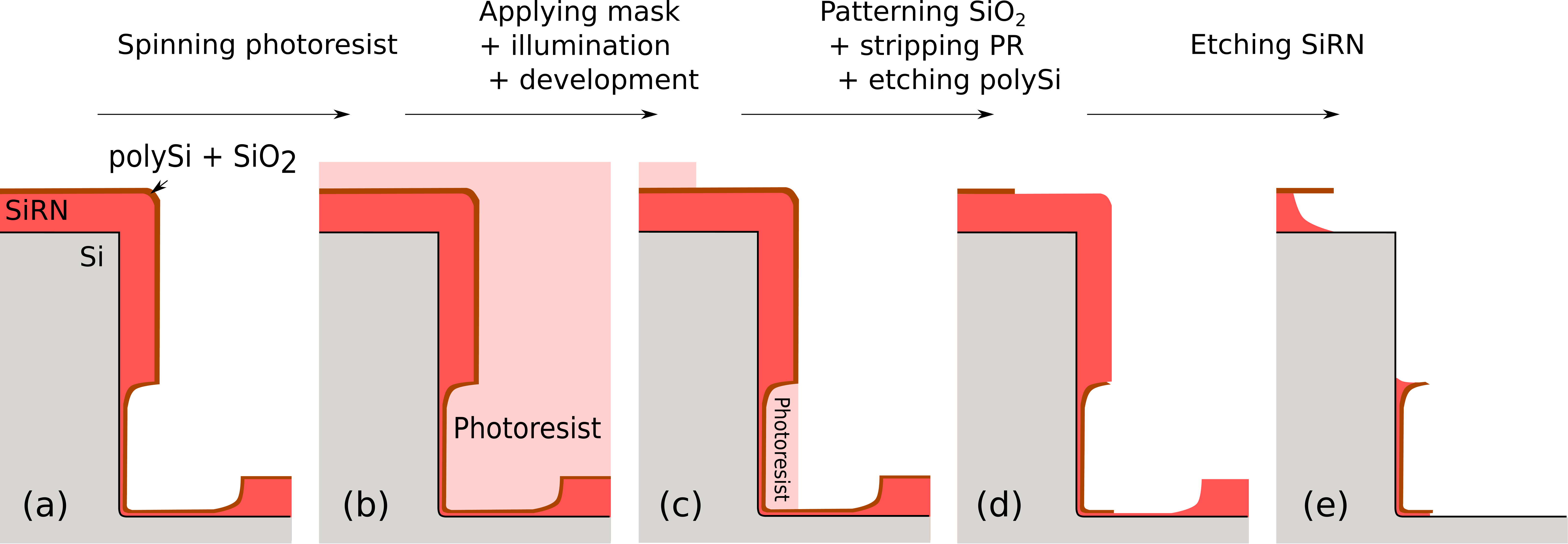}
\caption{Difficulties arising when implementing corner lithography in an upright mold.  We consider the case where the process flow depicted in Fig.~\ref{fig:Fab Outline corner litho} has been followed up to step (i) inclusive.  (a): Dry etching is impossible, a wet etching strategy must be considered using a masking layer.  A polySi layer is deposited followed by short oxidation.  (b): Photoresist is spun over the wafer.  The use of a thick photoresist dedicated to high-aspect ratio structures protection is necessary to protect the molds.  (c): Directional nature of UV illumination makes impossible the proper patterning of photoresist under the thick \SiN plate.  (d): Consequently, the masking layer is not etched away everywhere.  (e): \SiN is still present all around the molds at their bottoms.}
\label{fig:90_issues}
\end{figure}

After performing the corner lithography, it is necessary to pattern the \SiN features before releasing the structures, Fig.~\ref{fig:Fab Outline corner litho}~(i--j).  While reactive ion etching is a perfect option for \SI{70.6}{\degree} stop-programmable hinges, it is impossible to use it in the case of the vertical molds obtained by dry etching.  The difficulties arising with upright sidewalls are twofold, as shown in Fig.~\ref{fig:90_issues}.  One is the difficulty to etch several \si{\um} of material from the top.  And the other is the vertical thick \SiN plate of the complex hinge, see Fig.~\ref{fig:Principle}, that makes impossible a proper illumination of the photoresist for the subsequent lithography steps.  Therefore an alternative method must be considered.  

\begin{figure}
\centering
\includegraphics[width=1\linewidth]{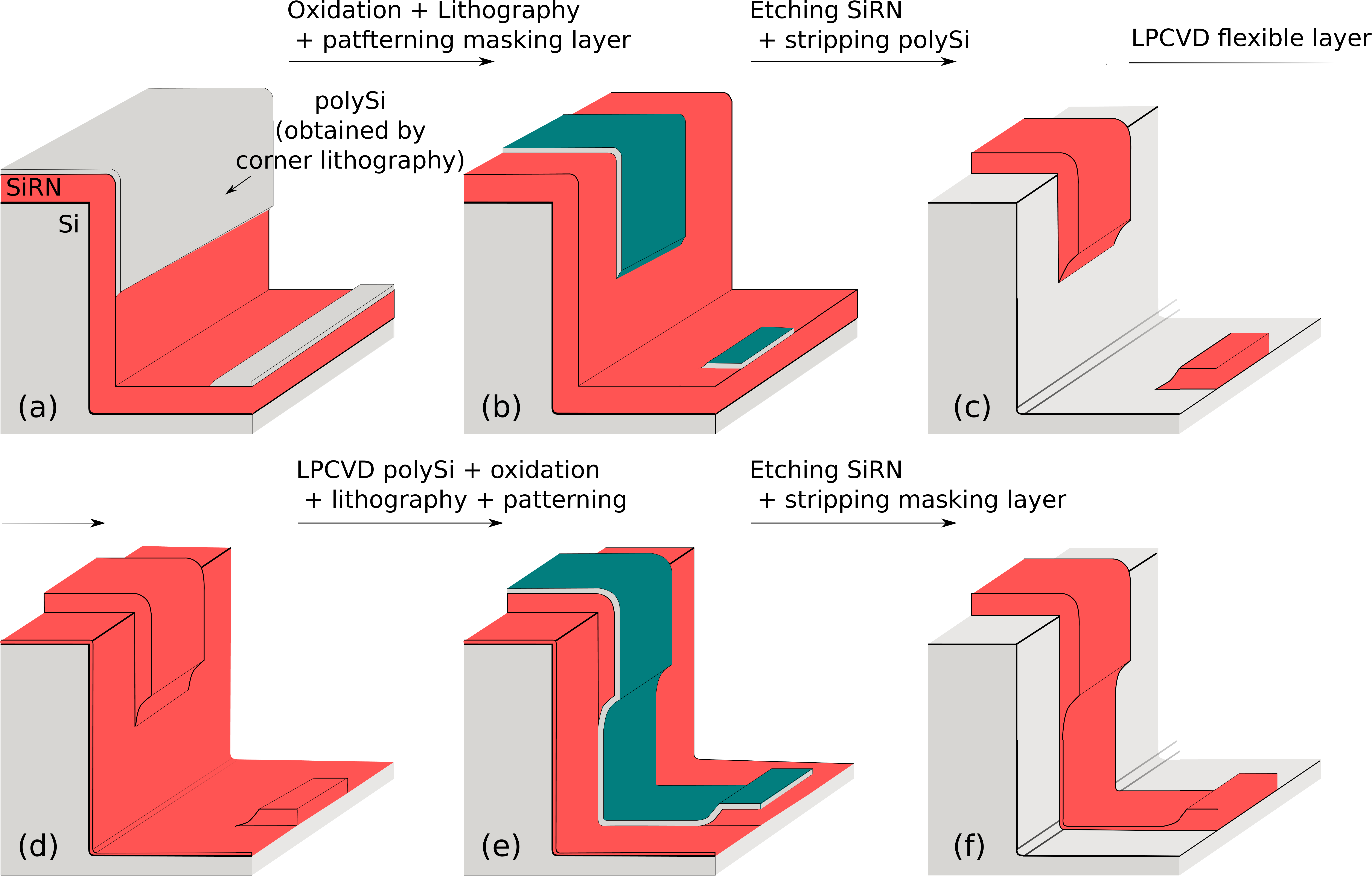}
\caption{ Alternative steps to pattern upright sidewalls.  (a): The thick \SiN should be patterned first.  The polySi masking layer obtained thanks to corner lithography can be used.  (b): A short oxidation step follows to form a thin \SiO layer.  Unlike in Fig.~\ref{fig:90_issues}, lithography is possible to pattern the masking layer.  (c): \SiN is wet etched using the oxidized polySi as a masking layer.  PolySi is fully oxidized and stripped.  (d): A flexible \SiN layer is deposited by LPCVD.  (e):  After LPCVD of a polySi layer followed by a short oxidation, the mask used in step (b) is applied a second time to pattern the masking layer.  (f) \SiN is wet etched and the masking layer stripped.   The origami patterns are now complete with no \SiN wire running anywhere around the mold as was the case in Fig.~\ref{fig:90_issues}.}
\label{fig:90_fab}
\end{figure}

We suggest here to pattern the vertical sidewalls in two main steps, which are shown in Fig.~\ref{fig:90_fab}.  Since it is impossible to illuminate the photoresist when it is masked by the thick upright part of the hinge, both parts of the hinges cannot be simultaneously patterned.  The whole thick \SiN plate must be patterned first by wet etching before depositing the flexible part.  In order to avoid an extra masking layer deposition and time-consuming wet etching step, the already patterned polySi layer by means of corner lithography, Fig.~\ref{fig:Fab Outline corner litho}-(h), can be used for this purpose.  A short oxidation step is performed to form a $\simeq 5~\si{\nm}$ \SiO layer on top of the polySi.  A lithography follows to define the overall geometry of the foldable objects.  \SiO is then patterned in three dimensions in wet etchant BHF.  After stripping the photoresist, TMAH is used to pattern the underlying polySi using the \SiO layer as a mask, Fig.~\ref{fig:90_fab}-(b).  The thick \SiN layer is then accessible and can be selectively etched, independently of the spatial direction.  

After the conformal deposition of a thin \SiN layer (150~\si{\nm}, Fig.~\ref{fig:90_fab}-(d)) followed by a polySi layer (100~\si{\nm}), the exact same procedure can be applied again: short oxidation of polySi,  second lithography using the same mask, patterning of oxide and polySi etching, Fig.~\ref{fig:90_fab}-(e).  Wet etching of thin \SiN and final stripping of masking polySi layer follow to complete the origami patterns, Fig.~\ref{fig:90_fab}-(f).  

The use of wet etching arranges that the \SiN layers are etched under the polySi masking layers.  The structures will therefore be attacked twice from their sides, in steps (c) and (f) in Fig.~\ref{fig:90_fab}.

\section{Results}

\subsection{Fabrication results---\SI{70.6}{\degree} stop-programmable hinges}

\begin{figure}
\centering
\includegraphics[width=.8\linewidth]{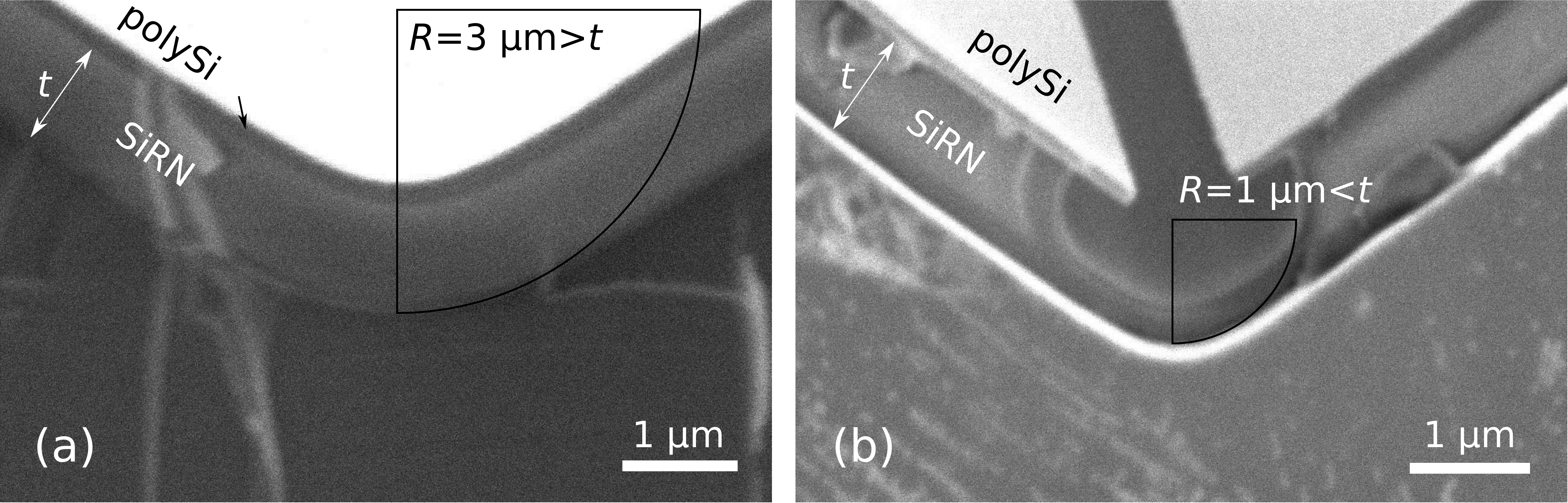}
\caption{Illustration of the design criterion described in Fig.~\ref{fig:Design_criterion}.  Figs. (a) and (b) show samples with different radiis of curvature for the initial mold after performing corner lithography and partially etching the bottom \SiN layer, Fig.~\ref{fig:Fab Outline corner litho}-(i).  The thickness $t$ of the first deposited \SiN layer is \SI{1.1}{\um}.  (a): When $R>t$, the entire surface of the polySi layer was oxidized  during step (f), consequently the retraction, step (g), had no effect.  (b): For $R<t$, the polySi layer is opened and the underlying \SiN layer can be etched, Fig.~\ref{fig:Fab Outline corner litho}-(i).}
\label{fig:706_radius_1_3}
\end{figure}

Fig.~\ref{fig:706_radius_1_3} illustrates the design criterion introduced in Fig.~\ref{fig:Design_criterion} and described in \Eref{eq:design_criterion}.  By varying the oxidation time, two molds with different radii of curvature were made (Fig.~\ref{fig:Fab Outline corner litho}-(b)) and subjected to the exact same fabrication steps until the isotropic etching of the bottom \SiN layer (Fig.~\ref{fig:Fab Outline corner litho}-(i)).  Since the thickness of the stack of the layers in the first case does not exceed $R$, there is no concave corner at the bottom of the groove.  Consequently, the timed-etching step of the top \SiN layer (Fig.~\ref{fig:Design_criterion}-(f)) does not yield nanowires and the entire polySi layer is oxidized during the subsequent step (g).  Without access points below the nanowires, no retraction can occur.  Once the oxide is stripped, a non-patterned thick \SiN and polySi layer stack is obtained, as can be seen in Fig.~\ref{fig:706_radius_1_3}-(a).  

In the correct case of Fig. (b), the radius of curvature $R$ is smaller than the thickness $t$ of the \SiN layer.  The entire etching procedure can proceed and yields a clear opening in the polySi layer through which \SiN can be etched.

\begin{figure}
\centering
\includegraphics[width=.8\linewidth]{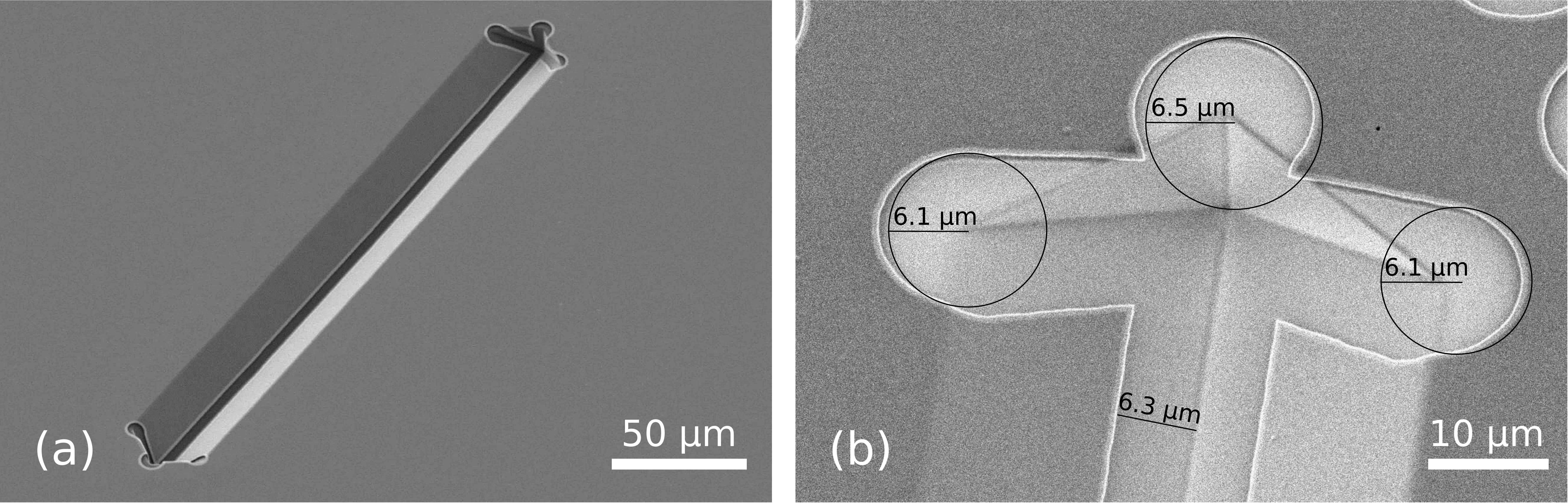}
\caption{(a): Overview of a V-groove after etching of the \SiN layer, Fig.~\ref{fig:Fab Outline corner litho}-(h).  (b): Zoom in the extremity of the groove after stripping away the polySi top layer, Fig.~\ref{fig:Fab Outline corner litho}-(i).  Retraction occurs in all concave corners, including the vertical planes.  The retraction length is the same in every direction.}
\label{fig:Retraction_groove}
\end{figure}

Corner lithography is a powerful three-dimensional patterning technique.  The trick works for concave corners of any size and spatial configuration.  As an illustration, Fig.~\ref{fig:Retraction_groove} shows the result when starting from V-grooves made by anisotropic etching in KOH.  The planes are organized in a well known fashion, and the use of corner lithography leads to retraction in all three dimensions.

\begin{figure}
\centering
\includegraphics[width=.8\linewidth]{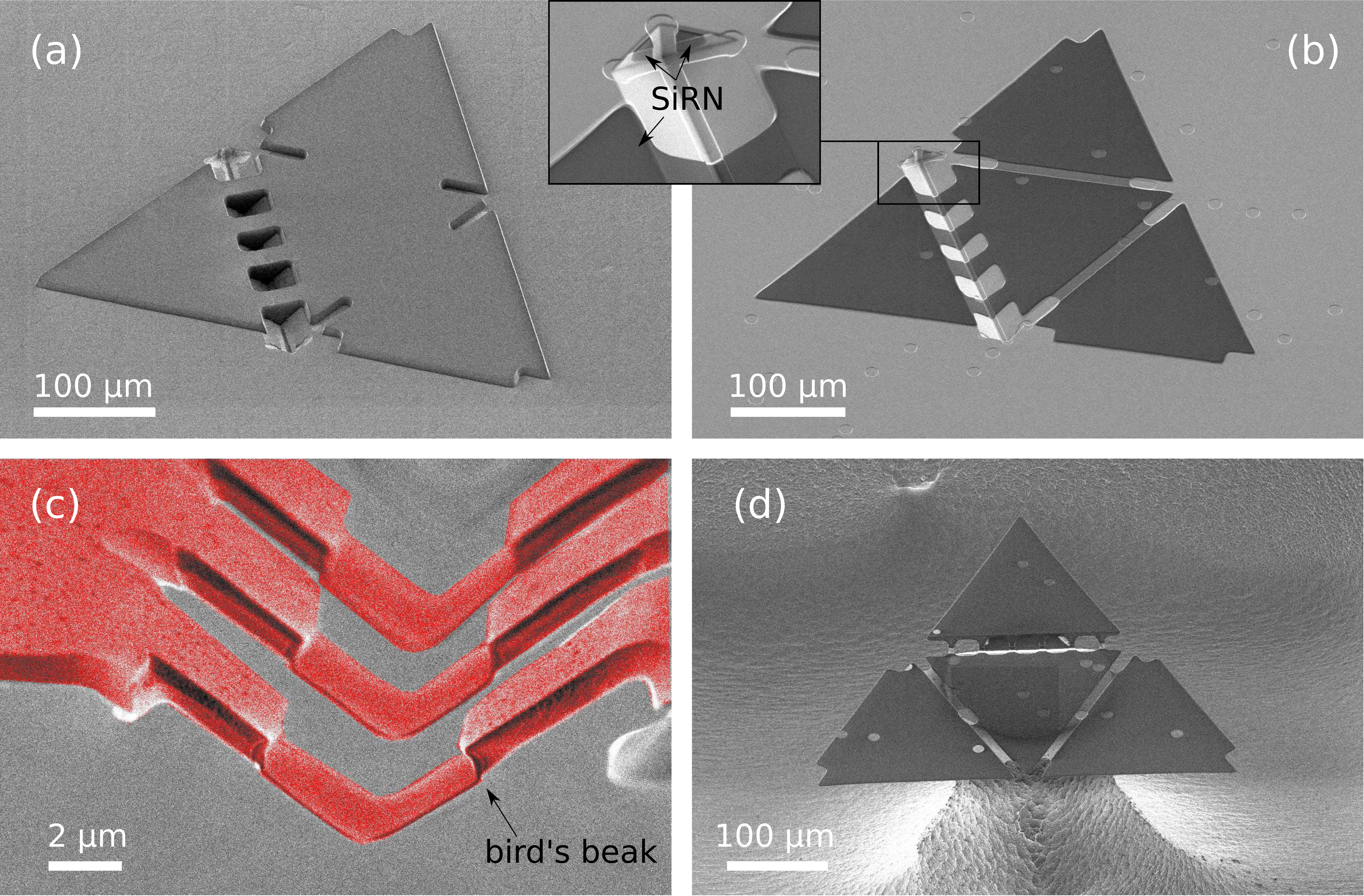}
\caption{SEM images of final fabrications steps.
(a): Tetrahedral pattern of thick photoresist.  (b): \SiN template after over-etching in the grooves and stripping of photoresist.  The combination of a complex hinge with two flat hinges can be observed, see Fig.~\ref{fig:pattern_tetra}-(a).  (c): Close-up image of a complex hinge after the silicon mold was etched away in dry etching and the photoresist stripped, Fig.~\ref{fig:Fab Outline corner litho}-(k).  (d): Overview of an unfolded tetrahedral structure at the end of the process.  Isotropic etching of Si is stopped once the pattern comes to rest on a central pillar, Fig.~\ref{fig:pattern_tetra}-(b).  Note that this structure is not completely released, since Si is still present under the flaps.}
\label{fig:706_last_steps}
\end{figure}

\cmtr{new fab details.}
Fig.~\ref{fig:706_last_steps} gives an overview of the last fabrication steps of a tetrahedral pattern.  The use of a thick photoresist developed for high aspect ratio features is necessary for good protection of the deep V-grooves during the dry etching of the \SiN features, Fig.~\ref{fig:Fab Outline corner litho}-(j).  The result obtained with AZ\textregistered~9260 photoresist is good, as shown in photographs (a) and (b).  Such planar protection is obtained by coating and spinning the resist at \SI{300}{rpm} for \SI{10}{\second} then \SI{60}{\second} at \SI{2400}{rpm}.  The resist is exposed three times for \SI{10}{\second} at intervals of \SI{10}{\second} and is developed for \SI{7}{\minute}.  As can be seen in the inset of picture (b), \SiN remains on the upright sidewalls after the dry etching step: only the top part was attacked by reactive ion etching.  These small \SiN spots are, however, not a problem for our folding structures.  

As is visible in pictures (a), (b) and (d), pinholes were present all over the wafer.  They originate from nanoscopic defects in the bottom \SiN layer that turn into microscopic features because of the corner lithography: another proof of the extreme sensitivity of the technique.  As long as the pinholes appear on the flaps and leave the hinges intact, they do not represent an issue for our folding purposes.  However, the quality of the first \SiN layer should be checked at the beginning of the fabrication.

The stop-programmable hinge shown in Fig.~\ref{fig:706_last_steps}-(c) is nearly identical to the schematic presented in the process flow, Fig.~\ref{fig:Fab Outline corner litho}-(k), except for the small bumps visible at the transitions between the thin and the thick parts of the smart hinges.  These transitions are called `bird's beaks' because of their characteristic shapes~\cite{Shankoff1980}, and originate from the oxidation step necessary to remove the polySi layer between steps (h) and (i) in Fig.~\ref{fig:Fab Outline corner litho}.

A nearly released tetrahedral pattern is presented in Fig.~\ref{fig:706_last_steps}-(d).  A circular protection of photoresist was present on top of it before the release step in SF$_6$, Fig.~\ref{fig:Fab Outline corner litho}-(k), hence the circular shape of the Si pillar.  

\subsection{Fabrication results---\SI{90}{\degree} stop-programmable hinges}

As was emphasized in Part~\ref{sec:90deg_fab}, the fabrication for both stop-programmable hinges is identical except for the creation of the molds and for the final etching of the \SiN layers.

\begin{figure}
\centering
\includegraphics[width=.8\linewidth]{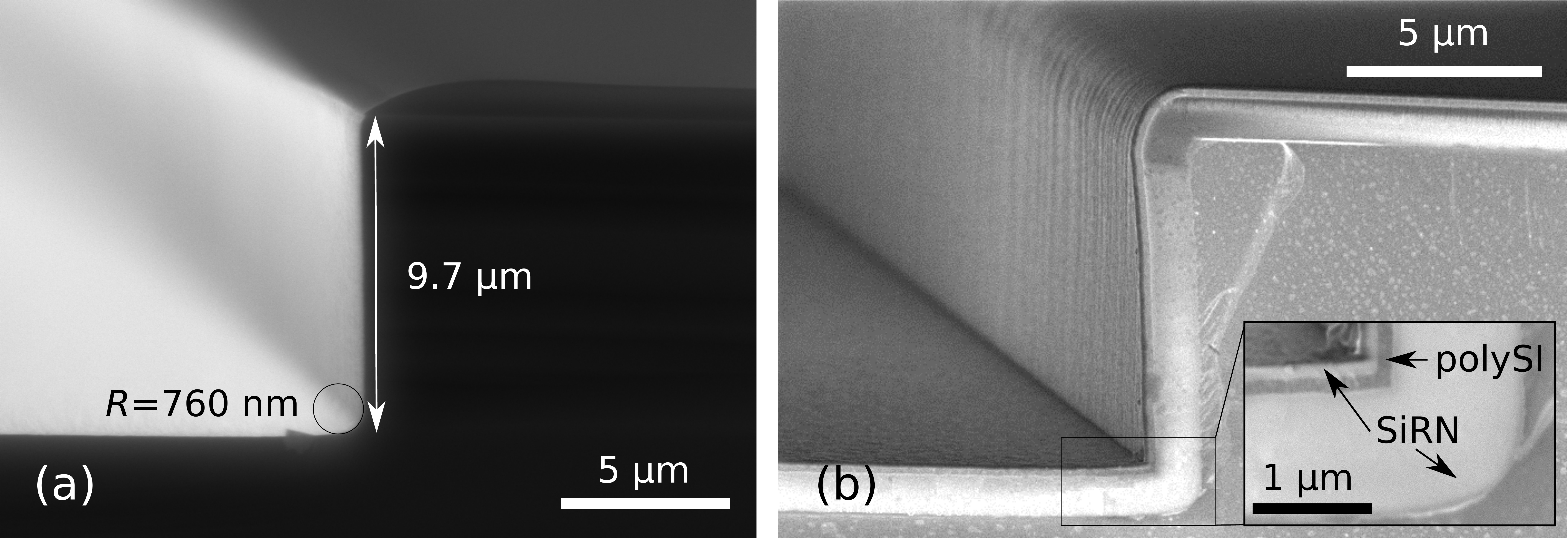}
\caption{(a): Cross section of a mold obtained
  by dry etching of silicon.  The corner is not exactly perpendicular (measured to be \SI{89}{\degree}) and the process  yields a round corner.  These parameters can be modified by fine tuning the dry etching step.  Note that photoresist is still present on top.  (b): Stack of the three layers necessary for corner lithography: thick \SiN (\SI{1070}{\nm}), polySi (around \SI{150}{\nm}) and a second layer of \SiN (\SI{146}{\nm}).}
\label{fig:cs_ASE}
\end{figure}

\cmtr{new details fab}
Our best results for getting molds using dry etching are presented in Fig.~\ref{fig:cs_ASE}.  We used a mixture of SF$_{6}$ (\SI{200}{sscm}) and O$_{2}$ (\SI{15}{sscm}) gases at \SI{-110}{\degreeCelsius} (Adixen AMS100 Reactive Ion Etcher, pressure \SI{1.6d-2}{mbar}, RF \SI{200}{\watt}, LF \SI{20}{\watt} on/off time \SI{25/75}{m\second}).  The dry etching step yields a rounded mold, picture (a), which reduces the oxidation time necessary to get the final desired radius of curvature.  The depth of the molds was checked on six different spots spread over one dummy wafer and was found to be the same as that depicted in Fig.~\ref{fig:cs_ASE} within an error of \SI{5}{\percent}.  Photograph~(b) shows a similar mold on top of which the three layers necessary for corner lithography were deposited, Fig.~\ref{fig:Fab Outline corner litho}-(d).  Since the design criterion (\Eref{eq:design_criterion}) is respected, the rounding has disappeared after the conformal deposition of the first \SiN layer.

In Fig.~\ref{fig:cs_ASE}-(b), the defects of the initial mask can be observed on the vertical sidewalls.  In order to avoid undesirable corner lithography starting points at that location, the over-etching of the top \SiN layer (Fig.~\ref{fig:Fab Outline corner litho}-(e)) was deliberately long.  The corners formed due to irregularities of the mask have large opening angles, so the surplus of material is thinner than in the \SI{90}{\degree} corners at the bottom of the grooves.  Since in a perpendicular corner the material is in theory \SI{41}{\percent} thicker (see Fig.\ref{fig:Principle corner litho}), a \SI{25}{\percent} over-etch was performed.

\begin{figure}
\centering
\includegraphics[width=.8\linewidth]{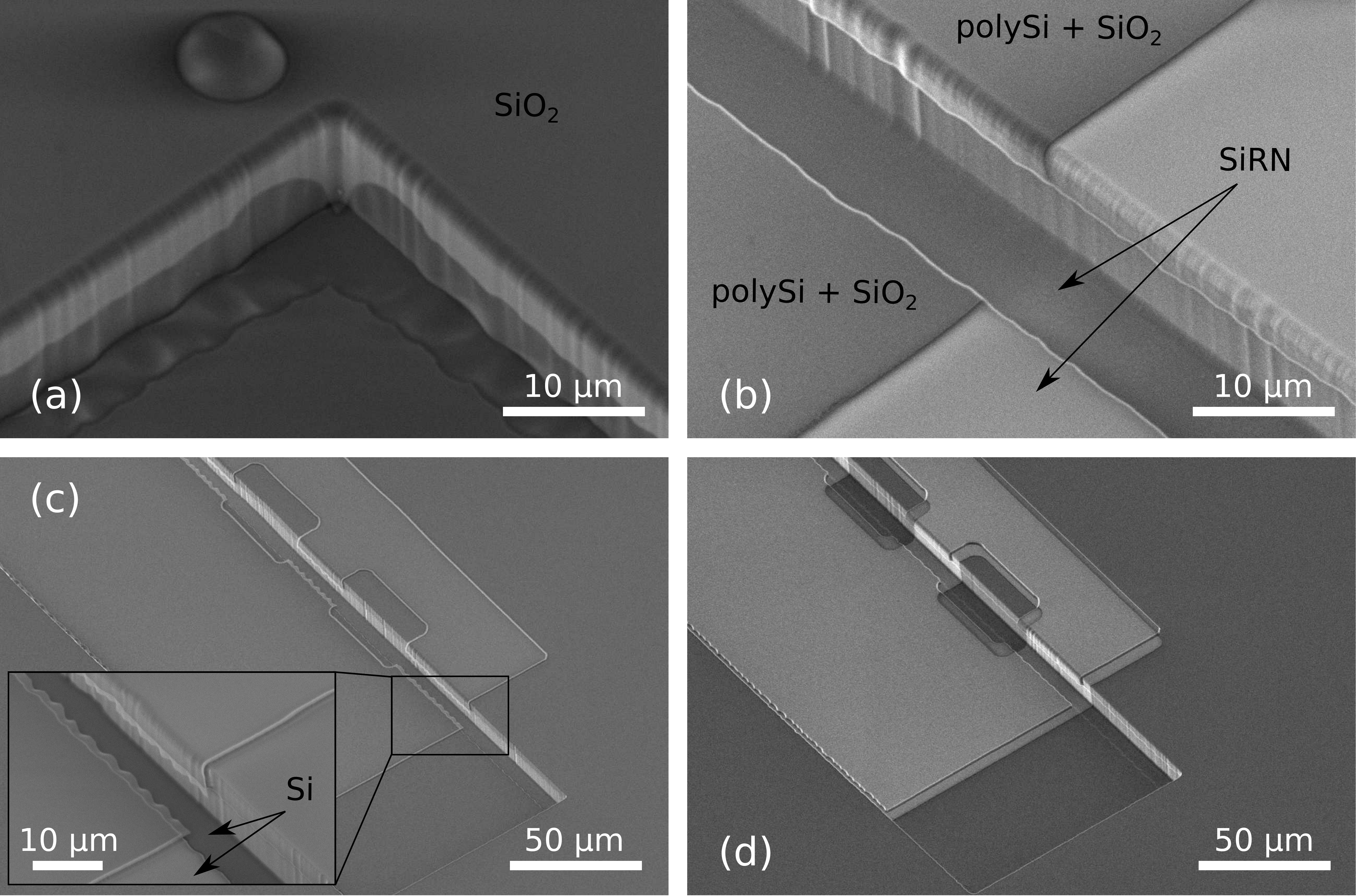}
\caption{SEM images of fabrication steps for \SI{90}{\degree} complex hinges.  (a): Retraction of polySi is visible under \SiO, Fig.~\ref{fig:Fab Outline corner litho}-(g).  The polySi plate is not exactly straight.  These irregularities are exact replicas of the defects of the silicon mold, which have been magnified through corner lithography.  (b): Thick \SiN with patterned masking partially oxidized polySi layer on top.  (c): Same as (b) right after wet etching of \SiN.  Stress in \SiO mask causes the curtain-like overhanging thin film.  (d): Final structure before release after the second wet etching procedure was applied.}
\label{fig:90_process_results}
\end{figure}

Fig.~\ref{fig:90_process_results} shows the subsequent steps in fabricating \SI{90}{\degree} stop-programmable hinges.  The successful retraction after the corner lithography is shown in photograph (a).  Unlike the results presented in Fig.\ref{fig:Retraction_groove}, no retraction occurred on the top or in  the corners of the mold.  This is a consequence of the long over-etching explained in the previous paragraph.  This long over-etch of the top \SiN layer when performing corner lithography removed all unwanted material in the irregularities of the molds, as well as in their corners.

As explained in Part~\ref{sec:90deg_fab}, dry etching cannot be used in the case of upright molds and a double wet etching strategy must be used.  Fig.~\ref{fig:90_process_results}-(b) shows the patterned masking layer, corresponding to step (b) in Fig.~\ref{fig:90_fab}.  The oxidized polySi layer following the retraction shown in (a) was further shaped by oxidation and subsequent lithography steps.  This way, the material at the bottom of the molds is etched and the geometry of folding patterns is defined simultaneously.  Picture (c) shows the resulting structures after \SiN etching, with only the flexible part of the hinges missing, see Fig.~\ref{fig:90_fab}-(c).  The same wet etching procedure is applied a second time after depositing a thin \SiN and polySi layer, resulting in the structures shown in photograph (d), Fig.~\ref{fig:90_fab}-(f).  Misalignment of the mask in the second lithography step on top of the first patterned \SiN explains the staircase-like shape of the final \SiN object.

\begin{figure}
\centering
\includegraphics[width=.8\linewidth]{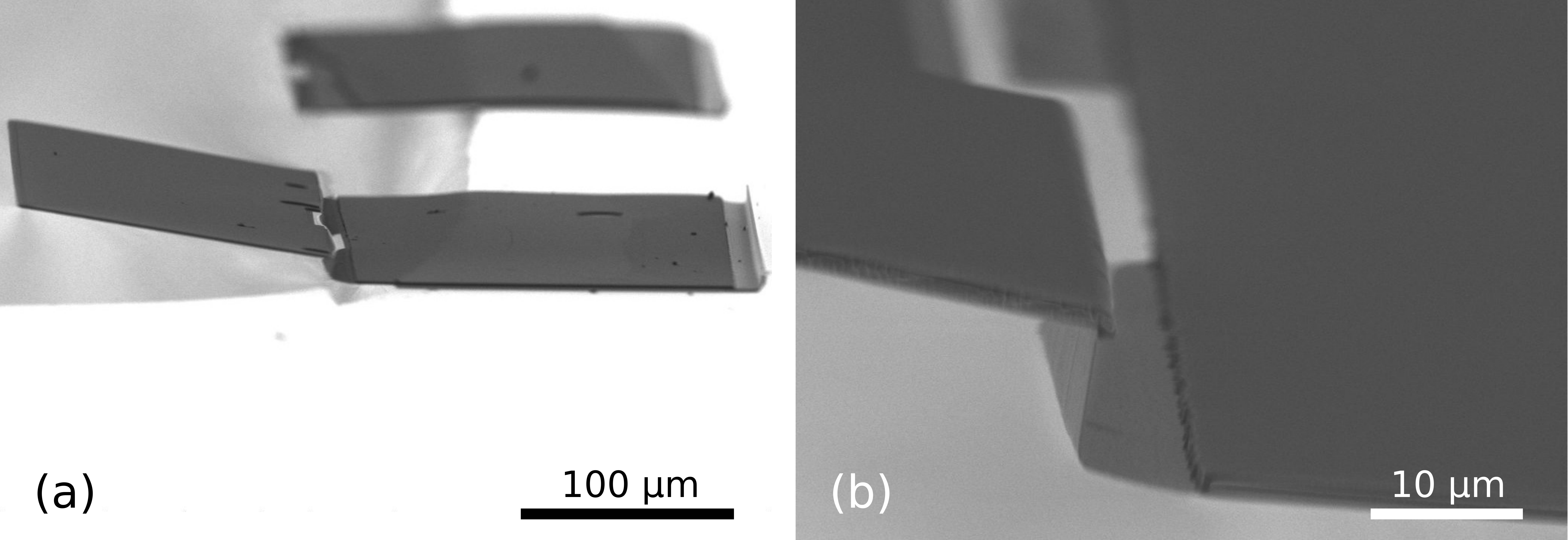}
\caption{(a): Final structure before folding.  Residual stress in the thin \SiN layers provoked a slight bending upward of the flap.  (b): Zoom in at a stop-programmable hinge.}
\label{fig:results_fab_90}
\end{figure}

Fig.~\ref{fig:results_fab_90} shows a self-foldable object released from the Si substrate, similar to the structure in Fig.~\ref{fig:90_process_results}-(d).  The fabrication was in principle successful, except for the fact that the length of the landing part of the complex hinge is extremely small in comparison with the flexible part, see Fig.~\ref{fig:Principle_sph}.

\subsection{Folding experiments}

Self-folding experiments were carried out by manually depositing water droplets of 5 to \SI{15}{nL}.  An accurate positioning system allowed us to deposit the liquid right at the centre of the templates using a hollow fibre (\SI{50}{\um} diameter) connected to a high precision  Hamilton glass syringe filled with ultra pure water that was manually actuated.  The folding of the structures typically takes around one minute, depending on the size of the structures and the volume of liquid deposited.

\begin{figure}
\centering
\includegraphics[width=.8\linewidth]{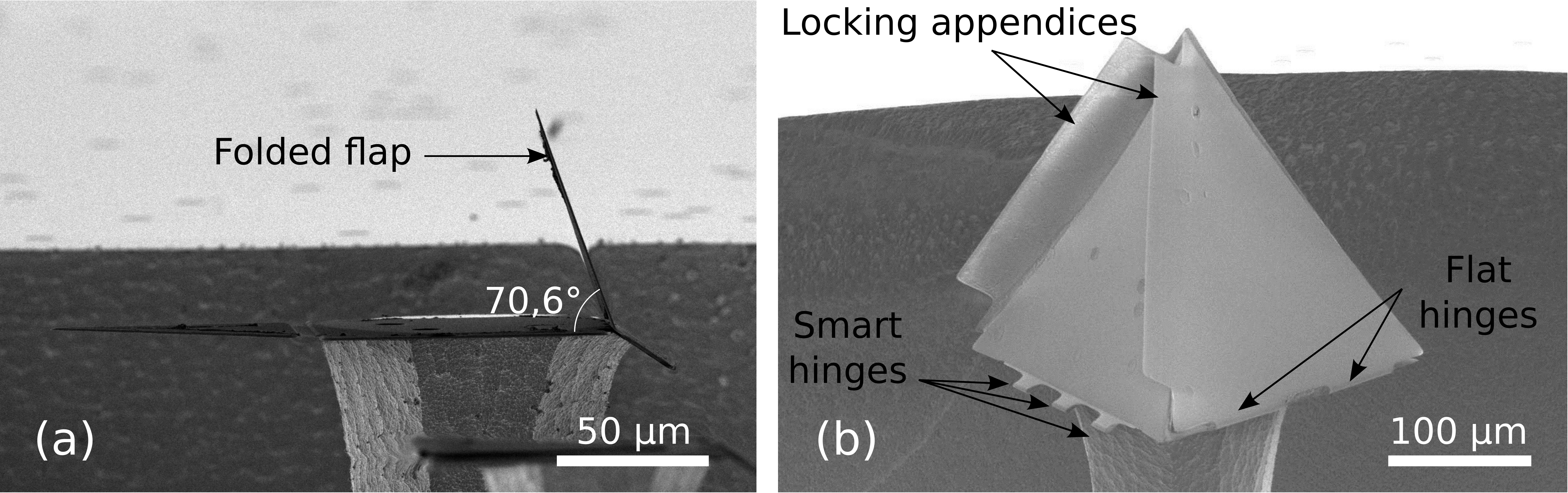}
\caption{Folded structure arrested by \SI{70.6}{\degree} stop-programmable hinges.  (a): Side view of an extruded-2D structure.  The flap is \SI{100}{\um} wide and the complex hinge has an original width of \SI{40}{\um}.  The flap of the left hand side, connected with a flat hinge, reopened after folding because of an insufficient bonding area.  (b): Folded tetrahedron.  The faces have sides of \SI{200}{\um}, the complex hinges are \SI{20}{\um} wide and the flat hinges are \SI{10}{\um} wide.}
\label{fig:70.6 folded structures}
\end{figure}

Fig.~\ref{fig:70.6 folded structures} shows the results of successful folding experiments.   Fig. (a) shows a  wing folded at the designed angle of \SI{70.6}{\degree}.  Fig. (b) shows a tetrahedron resting on a silicon pillar.  The large contact areas between the thick \SiN parts provide a good stability for the 3D objects after drying.  The tetrahedral structure was folded from a similar pattern presented in Fig.~\ref{fig:pattern_tetra}-(b) and Fig.~\ref{fig:706_last_steps}-(d).  One stop-programmable hinge makes sure that the folding stops at \SI{70.6}{\degree} while large appendices designed on the side of the other faces allow them to lock onto the first flap.

\begin{figure}
\centering
\includegraphics[width=.8\linewidth]{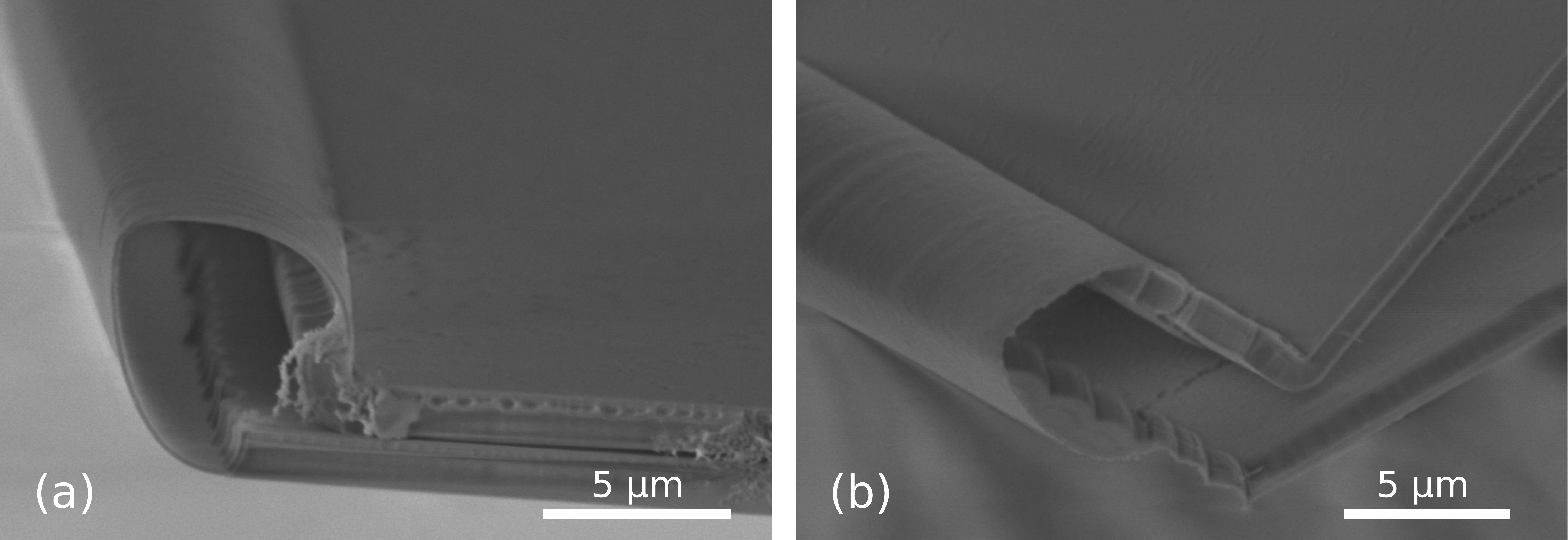}
\caption{Folding results of \SI{90}{\degree} stop-programmable hinges.  (a): The thick part of the complex hinge is too short (\SI{1.8(02)}{\um}) and fails to stop the folding.  Consequently, the flap is \SI{180}{\degree} folded.  (b): For this sample, the thick part is slightly longer (\SI{3.2(02)}{\um}).  This forces folding to stop at an intermediate position, but fails to stop the folding at \SI{90}{\degree} as was intended.}
\label{fig:90 folded structures}
\end{figure}

Fig.~\ref{fig:90 folded structures} shows the folding results when using \SI{90}{\degree} smart hinges presented in Fig.~\ref{fig:results_fab_90}.  These results are less successful.  A too long retraction of the polySi during the corner lithography, coupled with the wet etching steps necessary to pattern the final objects, led to wrongly shaped complex hinges.  The thick parts of the hinges are too small, and the folding did not stop at all (a), or stopped too late (b).  However, these results are encouraging and demonstrate that \SI{90}{\degree} stop-programmable hinges are feasible.  Reducing the TMAH retraction to \SI{2}{\um}, which would lead to \SI{4}{\um} long flexible plates, would most probably be sufficient for a successful assembly.

\section{Discussion}

The technique presented in this paper was first successfully used to fabricate \SI{70.6}{\degree} stop-programmable hinges.  The same principle was then applied to upright molds, but issues were encountered during the process.  Difficulties with the molds (retraction of the mask and roughness) caused our first attempt to fail.  However, it became straightforward to perform corner lithography in \SI{90}{\degree} molds when a long over-etching was used.  We initially used a 1.05 times over-etching in the case of \SI{70.6}{\degree} stop-programmable hinges and increased it to 1.25 for \SI{90}{\degree} complex hinges.  Later on during fabrication, difficulties in patterning  upright sidewalls forced us to modify the process.  Dry etching of \SiN with rotated samples, as well as a single wet etching step, were first unsuccessfully carried out before coming up with the appropriate two-step wet etching patterning strategy described here.  

 The results presented here are encouraging.  As long as the necessary molds can be obtained in Si, the procedure that has been described in this paper is applicable to virtually any folding angle.  In principle, the wet etching method for patterning \SiN onto the sidewalls should be applied for any mold opening angle $\alpha \le \SI{90}{\degree}$, while dry etching should be preferred when $\alpha > \SI{90}{\degree}$.

Elastocapillary folding allows the assembly of relatively large structures.  The characteristic capillary length, $\lambda_c =\sqrt{\nicefrac{\gamma}{\rho g}}$ (here, $\gamma$ is the surface tension, $\rho$ is the density of the fluid, and $g$ denotes the the acceleration due to gravitation), gives an indication of the scale on which capillarity is dominant over gravity.  For clean water and air at standard conditions, the transition is around \SI{2}{\mm}.   Elastocapillary folding of several \si{\mm} long silicon-based objects is therefore theoretically possible.  It is known to be hard to fabricate features of this size out of the wafer plane by conventional two-dimensional micro-fabrication techniques.  Indeed, \SI{2}{\mm} is more than four times the standard thickness of a standard silicon wafer.  \SI{90}{\degree} stop-programmable hinges are especially interesting since they would permit popping up several mm long features out of the plane of the silicon wafers exactly where the hinges are designed.  This technique, combining the strengths  of well known standard fabrication techniques with the ease of self-folding, could have many applications, such as 3D sensing, Micro-Opto-Electro-Mechanical Systems (MOEMS), or 3D memory.

Furthermore, we have shown that the use of corner lithography leads to retraction in all three dimensions (Fig.~\ref{fig:Retraction_groove}).  For the purpose of self-folding, only the features at the bottom of the molds were used, the others being etched away in Fig.~\ref{fig:Fab Outline corner litho}-(j).  Other features might be of interest for different purposes.  For example, unique 3D membranes or channels could be obtained.  Several applications have already been developed using corner lithography, such as the wafer scale fabrication of nano-apertures~\cite{Burouni2013,Berenschot2013}, photonic crystals~\cite{Yu2009} or micro-cages in which the culture of bovine cells has been demonstrated~\cite{Berenschot2012}.  

\section{Conclusion}

Starting from a simple micrometer-sized mask, accurate stop-programmable hinges for complex elastocapillary folding were micromachined in three dimensions using corner lithography.  In order to limit the stress in the hinges while folding, it was necessary to start with round molds.  Corner lithography  can be performed only if the combined thicknesses of the two first deposited layers is greater than the radius of curvature of the mold.  When this design criterion is respected, material can be accurately etched starting from any concave corner, independently of their spatial orientation.

The definition of the silicon molds at the start of the process determines both the locking angle and the accuracy of the stop-programmable hinges.  Selective KOH etching on (110) wafers yields well defined sharp corners that, in turn, can be used to micro-machine smart hinges that stop folding at  \SI{70.6(01)}{\degree}.  Micrometer sized three dimensional tetrahedral structures were successfully self-folded using capillary forces, thanks to these hinges.

We demonstrated the feasibility of \SI{90}{\degree} stop-programmable hinges.  Such complex hinges can be fabricated by making initial molds with dry etching at the cost of a poorer accuracy \SI{89(4)}{\degree} compared to wet etching.  Extra care must be taken when using dry etching to avoid any undesirable sharp corners, and a long over-etching during corner lithography must be performed to remove unwanted material.  Moreover, small irregularities in the molds will be magnified by the process and will cause defects in the final shape of the flaps.  

Stop-programmable hinges extend the possibilities for implementing the elasto-capillary folding of microstructures.  Using a simple filling procedure, millimeter-long silicon-based structures can be accurately popped out of the plane.  We believe that the accuracy and versatility of the technique will find widespread application in 3D sensing, MOEMS, or 3D electronics, for instance.

\section{Acknowledgements}
The authors would like to thank R.  G.  P.  Sanders for his valuable help with the folding experiments, K.  Ma for his precious help with the cleanroom work, and M.  J.  de Boer for his advice on dry etching.  This paper would have never been possible without J.  W.  van Honschoten, who inspired this project.

\section*{References}


\end{document}